# Software Engineering und Software Engineering Forschung im Zeitalter der Digitalisierung

Ein Beitrag über das aktuelle und zukünftige Selbstverständnis des Software Engineering

> "Die Informatik verändert sich. Deshalb müssen auch die Informatiker Ihre Rolle in der Gesellschaft neu überdenken."
> von Ranga Yogeshwar 2018 auf der Informatik Tagung der GI.

Michael Felderer, Universität Innsbruck
Ralf Reussner, KIT Karlsruhe
Bernhard Rumpe, RWTH Aachen

## Aktueller Stand und Beobachtungen

Die Digitalisierung wirkt sich nicht nur auf die Gesellschaft aus, sie verlangt auch eine Neubestimmung des Standorts der Informatik und der InformatikerInnen, wie sie der Wissenschaftsjournalist Yogeshwar in seinem obigen Zitat anregt. Da gerade sämtliche Aspekte der Digitalisierung auf Software beruhen, soll dieser Artikel der Versuch einer Neubestimmung insbesondere der Rolle des Software Engineering und seiner Forschung sein. Schon jetzt haben softwarebasierte Produkte, Systeme oder Dienstleistungen praktisch alle Lebensbereiche durchdrungen. Dadurch wird Software und damit das Software Engineering zum kritischen Baustein und zentralen Innovationstreiber der Digitalisierung in allen Lebensbereichen. Auch wissenschaftlich ergeben sich neue Chancen und Herausforderungen für das Software Engineering als treibende Disziplin bei der Entwicklung jedweder technischer Innovation. Gerade die Chancen dürfen allerdings auch nicht dem Wettbewerb um bibliometrische Zahlen als Selbstzweck geopfert werden.

### Trends

Konkret können wir gegenwärtig wenigstens diese sieben Trends über Software und das Software Engineering beobachten:

1. Software spielt in vielen Produkten die Rolle des wesentlichen Innovationstreibers oder Produkte werden durch softwarebasierte Dienstleistungen ergänzt, um sich am Markt wesentlich zu differenzieren. Beispiele sind automatisiertes Fahren oder die Digitalisierung von Geschäftsprozessen.

2. Software integriert Geräte und Dienstleistungen und ist damit der Kern sogenannten Systems-of-Systems. Beispiele sind die Themen zusammengefasst unter "Industrie 4.0", das Smart Energy Grid oder moderne multimodale Mobilitätssysteme.
3. Software wird sogar noch zunehmend von Personen entwickelt, die dies nicht primär gelernt haben. Das liegt zum Teil an der Unterversorgung des Arbeitsmarktes und zum Teil an dem zur Entwicklung notwendigen Domänenwissen. Notwendiger- und glücklicherweise wird zumindest die Programmierung von Software zum Commodity, deren Möglichkeiten und Grundzüge allerdings jeder wenigstens ansatzweise verstehen sollte.
4. Software nutzt zunehmend KI-Techniken, um Funktionalitäten zu lernen, zu verbessern oder anzupassen. Dadurch wird die Überprüfung von Qualitätseigenschaften während der Entwicklungszeit eingeschränkt und manchmal sogar verhindert, da das eigentliche Verhalten der Software erst nach der Trainingsphase absehbar ist, und nicht nur durch bloße Analyse des Codes oder das klassische Testen von Anforderungen deduziert werden kann. Beispiele finden sich etwa in der Bild- und Sprachverarbeitung und damit auch in sicherheitskritischen Systemen, wie autonomen Fahrzeugen. Obwohl dies zum Beispiel aufgrund der Nutzung von Datenbanken immer schon partiell so war, übernehmen nun die Daten eine stärkere Rolle bei der Definition des Verhaltens von Software. Zudem haben KI-Techniken (etwa durch neue Verfahren der Bild- und Spracherkennung) auch zu einer Erweiterung des Spektrums der Benutzerschnittstellen und damit der Bedeutung der User Experience beigetragen.
5. Auch für den Forschungsprozess im Software Engineering selbst sind Daten entscheidend, um Hypothesen empirisch zu prüfen. Allerdings birgt die beinahe beliebige Verfügbarkeit von Daten im Software Engineering auch Gefahren für den Forschungsprozess. Zu häufig werden Daten etwa aus Software-Repositorien oder durch Umfragen generiert und anschließend publiziert ohne diese durch eine Theorie, eine genaue Kontextdefinition oder eine Vision zur Entwicklung einer Methode zu fundieren, wodurch der Nutzen dieser Arbeiten für die Disziplin des Software Engineering unklar bleibt.
6. Software greift immer normativer in Lebensrealitäten ein. Software-Produkte haben bereits heute institutionellen Charakter: sie regeln und machen Vorgaben, allerdings oft ohne die Legitimation der üblichen Institutionen. Sie werden daher in Zukunft notwendigerweise zunehmend von staatlicher Seite reguliert. Dazu gehört insbesondere der Schutz personenbezogener Daten als essentielle Komponente des Schutzes der Personen selbst.
7. Der Umgang mit Software wirkt sich auch auf die Kognition und die Persönlichkeit der BenutzerInnen sowie auf die Gesellschaft als ganzes aus. Diese Wirkungen sind heute zwar erkennbar aber noch nicht wirklich abgeschätzt oder gar reflektiert bei der Entwicklung der Software.

Mit der Industrialisierung und der Vergesellschaftung von Software und Daten einhergehend ist auch zu beobachten, dass immer mehr Menschen aktiv in die Entwicklung von Software eingebunden sind, sei es durch die Bereitstellung von Anforderungen, das Testen von Systemen im Feld, oder die Teilnahme an der Softwareentwicklung selbst und dass mit immer kürzeren Innovationszyklen auch immer kürzere Entwicklungszyklen erfordern.

Konkret wird Software zur Laufzeit durch nachladbare Komponenten und Apps stark veränderlich, erweiter- und anpassbar. Der Mensch ist als Datenquelle heute (mehr passiv) Teil von Software-Ökosystemen. Aber auch zur aktiven Konfiguration (z.B. der vielen einzelnen Geräte im zukünftigen E-Home) werden immer mehr Menschen die Fähigkeit benötigen, rudimentär Software zu programmieren ("Programming as a Commodity"). Zudem rückt der Mensch als User immer mehr in den Fokus der Entwicklung, um durch neue Interfaces wie Sprachein- und Ausgabe oder Augmented Reality die User Experience zu optimieren. Software Engineering muss sich also in Zukunft nicht nur mit der Entwicklung von Software beschäftigen, sondern auch mit deren hoch-adaptiver Konfigurierbarkeit, Benutzerakzeptanz und der Qualitätssicherung von konfigurierter und durch Daten trainierter Software.

Schon aus technischer Perspektive ergeben sich aus den obigen Beobachtungen neue Herausforderungen, die auch eine neue Standortbestimmung der nunmehr 50 Jahre alten Disziplin des Software Engineerings [1,2,3,4], erfordern:

(1)     Software Engineering nimmt Regulationen und Werteverständnisse auf und versucht Verfahren, Algorithmen und deren Konfigurationsdaten sowie Architekturen zu entwickeln, um diese systematisch umzusetzen.

(2)     Software Engineering verpackt Software-Ansätze so, dass auch Nicht-SoftwaretechnikerInnen sie nutzen können. Als Beispiel und Vorbild können Datenbanksysteme gelten, die auch von Nicht-InformatikerInnen sinnvoll eingesetzt werden können. Die Nutzbarmachung wesentlicher Algorithmen, insbesondere des maschinellen Lernens ist für andere Disziplinen ist eine weitere Schnittstelle, zugegebenermaßen nicht nur, aber auch zu den Sozialen- und Gesellschaftswissenschaften.

(3)     Software Engineering beschäftigt sich natürlich primär mit Software und den notwendigen Grundlagen ihrer Entwicklung. Aber auf beide Bereiche haben Menschen und Organisationen einen entscheidenden Einfluss. Die Entwicklung und Untersuchung von Software Engineering-Artefakten bedarf deshalb die Einbeziehung des Entstehungs- und des Anwendungskontextes. Unter anderem deshalb ist die Anwendung von empirischen Forschungsmethoden und Theorien aus den Sozialwissenschaften wie Psychologie oder Soziologie notwendig, um menschliche und organisatorische Kontextfaktoren geeignet in den Software Engineering Forschungsprozess weiter zu integrieren. Software Engineering hat in diesem Sinne nicht nur Züge von traditionellen Ingenieurwissenschaften wie Maschinenbau oder Elektrotechnik, sondern unterscheidet sich von diesen durch diesen zusätzlichen sozialwissenschaftlichen Charakter.

## Konsequenzen für das Software Engineering

Daraus lassen sich einige Konsequenzen für Software Engineering ableiten:

- (1) Eine der wesentlichsten Konsequenzen aus diesen Beobachtungen ist, dass Software bzw. die Handlungen, Vorschläge und Entscheidungen die Software trifft oder ausführt in Zukunft besser nachvollziehbar sein müssen. Dies gilt insbesondere bei adaptiver Software, bei lernender Software und bei Software, die in dem oben genannten Maß durch EndnutzerInnen modifizierbar sein soll. Solche Software zu erstellen ist natürlich auch für Software Engineering eine Herausforderung, da sowohl die Form der Erklärung für NutzerInnen einfach und gut verstehbar sein muss, als auch zum Beispiel bei adaptiver und lernender Software Erklärungen nicht notwendigerweise vorab bereits definiert und fixiert werden können.

(2) Software Engineering drängt sich vermehrt in die Rolle des umfassenderen Systems Engineerings. Durch die zunehmend dominierende Rolle von Software in technischen Systemen verändert sich der Zusammenhang von Software und Systems Engineering. Konnte man Software Engineering bisher als einen Teil des Systems Engineering betrachten, so bietet heute Software Engineering selbst oft die grundlegenden Methoden für die Entwicklung softwareintensiver technischer Systeme. Software Engineering entwickelt sich deshalb von einer "Nebenbeschäftigung" des Systems Engineers zu dem primären Treiber der Entwicklung. Daraus lassen sich einige Konsequenzen ableiten, die vor allem Entwicklungsprozesse betreffen. Immer noch beißen sich der funktionsorientierte Softwareentwicklungsprozess mit dem vor allem geometrisch dekomponierenden Systementwicklungsprozess. Viele Methoden zur Systematisierung und Automatisierung der Software Engineering Aktivitäten bei Dekomposition, Synthese, Qualitätssicherung aber auch im Management von Evolution und Varianten müssen noch in den Systementwicklungsprozess übertragen werden. Erst dann lassen sich auch die von vielen gewünschten agilen Vorgehensmethoden stärker und gewinnbringend in die Systementwicklung einbringen.

(3) Software Engineering war es immer und wird es noch stärker werden: eine Integrationsdisziplin. Durch Software werden einzelne Produkte und Dienstleistungen vernetzt. Damit ergibt sich durch Software oft erst die Möglichkeit, innovative Szenarien umzusetzen, z.B. für Energieversorgung mit regenerativen Energieformen, Mobilität in der Kombination verschiedener Verkehrsträger oder zur komplexen Wertschöpfungsketten vernetzte Produktionsanlagen. Dadurch kommt dem Software-Ingenieur und der Software-Ingenieurin aber auch eine wesentliche Rolle in der "Findung" oder sogar „Erfindung" und Definition der eigentlichen Funktionalität des Systems-of-Systems zu. Während sich durch die Vernetzung die Produkte immer mehr integrieren, differenzieren sich die Rollen und Fähigkeiten der an einem Produkt beteiligten EntwicklerInnen immer weiter aus. Dies gilt sowohl für Produkte, als auch für die immer stärker vernetzte und digitalisierte Welt der Produktion. Dabei nicht zu vernachlässigen sind auch die immer komplexeren Softwarewerkzeuge, die zur Entwicklung selbst eingesetzt werden und aufgrund der Vernetzung ihrerseits die Verbindung zum ausgelieferten, aber beobachtbaren und aktualisierbaren Produkt halten werden.

(4) Software wird daten-getrieben adaptiv. Durch den Einsatz von Verfahren des Maschinellen Lernens zur Erbringung von Software-Funktionalität kann diese häufig nicht mehr zur Entwicklungszeit überprüft werden, da die Funktionalität nicht nur durch den Code, sondern auch durch (Trainings-)daten festgelegt wird. Hier müssen wir verstehen lernen, unter welchen Bedingungen und inwieweit ein zu bauendes System noch zur Entwicklungszeit absicherbar ist, unter welchen Bedingungen eine Laufzeit-Überwachung durch sogenannte „Monitore" möglich ist, wann eine a-posteriori-Analyse von Fehlern noch zulässig ist oder auch wann ein System eben so nicht gebaut werden darf, da es nicht mehr absicherbar ist. Und dies gilt sowohl dann, wenn das Training selbst bereits in der Entwicklungszeit abgeschlossen wird, als auch dann, wenn das System in Betrieb weiter lernt ("life-long-learning").

(5) Software muss normen- und gesetzes-sicher gebaut werden. Durch die notwendige zunehmende Regulierung software-basierter Produkte und Dienstleistungen gerade im Bereich des Datenschutzes und der Personensicherheit ergeben sich neue Anforderungen an die Software und ihre Entwicklungsmethodik. Bisher können Anforderungen wie der Schutz personenbezogener Daten gemäß sich wechselnder Einverständniserklärungen für verschiedene sich ändernde Arten der Datennutzung über den gesamten Lebenszyklus der Software nicht systematisch umgesetzt geschweige denn nachgewiesen werden.

Gemeinsam ist all diesen Herausforderungen, dass zum einen dadurch die Komplexität der Software weiter massiv ansteigen wird, zum anderen, dass immer mehr Menschen in verschiedenen Rollen mit Software und ihrer Entwicklung zu tun haben werden und von ihrer Funktionalität und von ihren Entscheidungen noch stärker als bisher abhängen werden. Insbesondere ist eine oben genannte Konsequenz daraus, dass auch der Bedarf an Nachvollziehbarkeit und Erklärbarkeit von Software steigen wird. Nur so wird auch die notwendige Akzeptanz beim Nutzer, bei Organisationen und in der Gesellschaft insgesamt erreichbar sein. Auch ist klar, dass es noch weniger als heute Personen geben wird, die ein Gesamtverständnis - oder auch nur einen detaillierten Überblick - haben werden. Die Konsistenzhaltung der verschiedenen Sichten auf ein System wird also immer wichtiger.

## Weiterentwicklung des Software Engineering

Im Zuge des Jubiläums 50 Jahre Software Engineering [1]  - seit der NATO Konferenz 1968 -  gibt es mehrere Diskussionen über deren Essenz und weitere Entwicklung des Software Engineering als Disziplin. Basierend auf der hohen praktischen Relevanz von Software Engineering Lösungen schlagen Basili und Briand [2] eine stärkere Kontextorientierung der Software Engineering Forschung vor, da die Anwendbarkeit und Skalierbarkeit von Software Engineering Lösungen zentral von Kontextfaktoren wie dem Menschen (etwa seines Persönlichkeitstyps und Wissenshintergrunds), der Organisation (etwa Kosten- und Zeitrestriktionen) oder der Domäne (wie etwa der Kritikalität oder der geforderten Compliance zu Standards) abhängt. In ihrem kontextgetriebenen Forschungsparadigma für das Software Engineering schlagen sie demnach vor, nicht mehr weiter primär Top-down allgemeine Ansätze ohne Berücksichtigung von Kontextfaktoren zu entwickeln, sondern Bottom-up in Kollaboration mit Praktikern innovative Lösungen zu entwickeln und dabei Zusammenhänge zu erkennen, Theorien zu entwickeln und wissenschaftlich vor allem durch den Einsatz empirischer Methoden zu evaluieren. Damit in Verbindung stehend hebt Broy [3] den integrativen Charakter des Software Engineering für alle etablierten Ingenieursdisziplinen hervor sowie die Notwendigkeit die Software Engineering Ausbildung aufgrund der strategischen und gesellschaftlichen Bedeutung von Software dahingehend zu erweitern, dass Software-IngenieurInnen kompetent Entscheidungen von strategischer Tragweite treffen können. Booch [4] sieht Abstraktion, Separation of Concerns, Verteilung von Verantwortlichkeiten sowie Einfachheit zum Management von Komplexität als zentrale Fundamente des Software Engineerings, die es zukünftig etwa auf KI-Systeme anzuwenden gilt, um Antworten auf zentrale Fragen in diesen Systemen, wie etwa nach ihrem Lebenszyklus, ihre Qualitätssicherung oder ihre Konfiguration, zufriedenstellend beantworten zu können.

Wir denken, die Weiterentwicklung des Software Engineering und insbesondere die integrative Anwendung des Software Engineering in Kombination mit anderen Disziplinen ist in den nächsten Jahren essentiell. In den letzten 50 Jahren hat sich innerhalb des Software Engineering Wesentliches getan. Eine solide und große Sammlung an Methoden erlaubt uns, punktgenau Agilität mit der konstruktiven Entwicklung und Synthese der Software und der analytischen Qualitätssicherung zu verzahnen. Requirements Engineering ist seinem Wesen nach schon immer eine hochgradig an Innovationen interessierte Methodik. Software Architektur erlaubt es, funktionsorientiert, aufgabenorientiert, orientiert an den physikalischen Gegebenheiten oder datenorientiert komplexe Herausforderungen zu zerlegen, zu organisieren und in robuste Lösungen umzusetzen. Modellbildung hat als Wesenszug zur Abstraktion und Darstellung der funktionalen Essenz eines Systems in der Softwareentwicklung unter anderem durch die Standards UML und SysML, aber auch durch domänenspezifische Sprachentwicklungen Einzug gehalten. Varianten- und Versionsmanagement befinden sich genauso in der Werkzeugkiste des Software Ingenieurs und der Ingenieurin wie Simulations-, Testautomatisierungs- und Analysewerkzeuge zur Qualitätssicherung sowie Messwerkzeuge verschiedenster Arten.

Eine der großen Herausforderungen für Software Engineering-Forschungsergebnisse selbst bleibt allerdings ihre praktikable Umsetzbarkeit durch Werkzeuge. Werkzeuge entstehen häufig als Nebenprodukt wissenschaftlicher Forschung oder werden durch eine gemeinsame Community als Open Source Werkzeuge so weit vorangetrieben, wie es dieser Community möglich ist. Werkzeuge sind heute hochkomplex und benötigen sowohl einen guten Bedienungskomfort, als auch die relevante Funktionssicherheit und müssen auf große Projekte skalieren.

# Was hat Software Engineering auch für andere Disziplinen zu bieten

Software Engineering ist zunächst wie jede Ingenieursdisziplin eine konstruktive Wissenschaft. In den 50 Jahren seit Bestehen der Disziplin wurde zunehmend gelernt, welche Konsequenzen sich aus der Besonderheit von Software ableiten, nicht materiell zu sein. Aktuell ist Software eben dabei den Alltag von Individuen, Organisationen und letztlich der Gesellschaft in jeder Hinsicht so schnell wie kaum eine andere Disziplin zuvor zu verändern. Nahezu alle Disziplinen und Domänen sind aktuell dabei sich zu digitalisieren. Wir gruppieren dabei die Standortbestimmung des Software Engineerings in Beziehung zu den anderen Ingenieurswissenschaften, zu den Sozialwissenschaften und zu den Naturwissenschaften.

## Bezug zu anderen Ingenieurswissenschaften

Wie bereits einleitend beschrieben wird der Bezug zu anderen Ingenieurswissenschaften gegenwärtig geprägt durch

(a) den steigenden Anteil von Software in vielen technischen Systemen,

(b) der Vernetzung dieser Systeme zu Systems-of-Systems durch Software,

(c) dem hohen Anteil an Software beim Innovationsgrad technischer Systeme sowie

(d) dem antizipierten Nutzen, wenn spezifisch software-technische Vorgehensweisen auch im Systems Engineering angewandt werden.

Letztlich bedingen diese Faktoren einander, so dass wir sie gemeinsam mit dem letzten beginnend ausführen.

Im Vergleich zu den Artefakten anderer Ingenieursdisziplinen ermöglicht die Eigenschaft von Software, nicht-materiell zu sein, einen größeren Entwurfsraum, der kaum durch physische Materialeigenschaften eingeschränkt ist. Werkzeuge und Methoden sind unter dem Stichwort Agilität in den letzten Jahren intensiv weiterentwickelt worden, sodass schnelle Feedback- und Innovationszyklen in der Software üblich sind. Seit etwa 2000 wissen wir, dass die Time-To-Market im Internetzeitalter noch viel wichtiger geworden ist und im Internet "ein Jahr nur drei Monate" dauert. Negative Auswüchse des schnellen Entwickelns, wie etwa der Bananensoftware, die erst beim Kunden reift, sind durch methodische Maßnahmen, wie etwa aktive Beta-Tester, testgetriebene Entwicklung, frühe Reviews (z.B. auch Pair-Programming) oder kontinuierliches Monitoring des Kundenverhaltens und damit auch der Kundenzufriedenheit abgefangen worden.

Wenn Software selbst keine physikalische Manifestation hat, ist ihre "Produktion" reduziert auf den Build-Prozess, das Deployment und die Konfiguration. Da diese Vorgänge weitgehend automatisierbar sind, ist vor allem der Entwicklungsprozess, um den sich traditionell ja das Software Engineering kümmert, der limitierende Faktor. Dabei spielen die einzelnen entwickelnden Menschen, die Team-Struktur und -Zusammensetzung sowie die Randbedingungen durch die Organisation eine größere Rolle als bei anderen Ingenieurwissenschaften. Sie prägen den Prozess der Produktentwicklung erheblich und beeinflussen damit letztlich auch Qualitätseigenschaften des Produkts selbst intensiv. Diese menschlichen und organisatorischen Faktoren sind deshalb ebenfalls ein wichtiger Einflussfaktor und damit Untersuchungsgegenstand des Software Engineering. Darüber hinaus bietet die Software Engineering Disziplin hier eine Fülle von Techniken und Methoden, die dieses Zusammenspiel zwischen Menschen in seiner Effizienz und seinen Ergebnissen optimieren.

Dadurch ergeben sich für die Entwicklung softwareintensiver technischer Systeme eine Reihe von Möglichkeiten, Einsichten aus dem Software Engineering zu übertragen. Dazu gehören zum Beispiel Vorgehensmodelle, die nicht mehr von der Voraus-Planbarkeit eines ganzen Systems ausgehen, sondern eine dezentrale Entwicklung von Systemen erlauben, die lose gekoppelt sind und keinem gemeinsamen Entwicklungsprozess unterliegen, obwohl sie System-Komponenten und -Schnittstellen teilen. Dazu gehören zum Beispiel auch agile Verfahren zur Unterstützung schneller Innovationszyklen. Da Software letztlich auch ein Modell der Realität mehr oder weniger explizit und ausführlich in sich trägt, sind Verfahren der Modellierung aus dem Software Engineering, wie zum Beispiel die UML, SysML oder Meta-Modellierung, und die Synthese, Analyse und Transformation von Modellen von besonderem Interesse für das Systems Engineering, zumal der Umgang mit Modellen und deren Analyse und Simulation ohnehin zum bestehenden Ingenieursvorgehen passt. In der Praxis zeigt sich aber, dass die Modelle der Software Engineering und des Systems Engineering bzw. die zugrunde liegenden Paradigmen nicht einfach vereint werden können und daher noch viel grundlegende Forschung notwendig ist.

Besonders sichtbar ist die Nutzung von Modellierungstechniken, wenn diese nicht nur zur Entwicklung, sondern auch beim Betrieb der Software im Einsatz sind. Traditionell sind das zum Beispiel explizite Datenmodelle oder Geschäftsprozessmodelle, die manchmal in Grenzen aber öfter auch flexibel anpassbar sein sollen. In der Produktwelt sind das Digitale Twins, also rein virtuelle, durch Modelle und Simulationen betriebene „Zwillinge" physischer Artefakte, die auch Meta-Information enthalten und durch prädiktive Techniken zur Vorhersage von Eigenschaften aber insbesondere letztlich zur Kontrolle und Steuerung ihrer physischen Zwillinge eingesetzt werden können. Die gemeinsame, weil eng verzahnte Entwicklung des physischen Produkts und seines virtuellen Zwillings erfordert neue Techniken. Dies ist auch deshalb notwendig, weil auch viele physische Produkte mit der Zeit ergänzt, erweitert, aktualisiert oder durch Ersatzteile modifiziert werden und ein digitaler Zwilling dies reflektieren und idealerweise sogar antizipieren muss. Hier wird das bestehende Verständnis von „Digital Engineering" substantiell erweitert, in dem nicht „nur" bestehende Entwicklungsprozesse durch Software unterstützt werden, sondern nun auch softwaretechnische Methoden neue Entwicklungsprozesse ermöglichen. Eine Vision wäre die integrierte Modellierung und Analyse in einem gemeinsamen Entwurfsraum, im dem Software aber auch physische Artefakte methodisch gemeinsam behandelt werden. Natürlich müssen diese Verfahren systematisch mit Fragen der unterschiedlichen Lebenszyklen von Software und physischen Artefakten umgehen und dabei ihre Konsistenz sicherstellen.

## Bezug zu den Sozialwissenschaften

Es ist wie immer im Software Engineering: die Einführung neuer Systeme verändert die Verhaltensweisen und Aktivitäten (Geschäftsmodelle und -prozesse, Support, etc.) der NutzerInnen. Das muss so sein, denn sonst wäre die Software sinnlos. Die zurzeit massiv voranschreitende Digitalisierung wird daher starke Konsequenzen für Individuen, Organisationen, Unternehmen und die Gesellschaft haben, die durchaus mit denen der Industrialisierung im 19. Jahrhundert vergleichbar sind. Offensichtlich hat sich das Kommunikationsverhalten bereits massiv verändert. Untersuchungen zeigen aber auch

Änderungen der menschlichen Kognition durch die allgegenwärtige Nutzbarkeit digitaler Geräte oder der Arbeitsweisen von Teams. Die Auswirkungen neu entstehender Geschäftsmodelle und ihrer gesellschaftlichen Auswirkungen können im sozialen und im Arbeitsleben immens sein, wenn die die Wertschöpfung durch Automatisierung und dem damit auch verbundenen Wegfall traditioneller Arbeitsplätze zugunsten neu geschaffener digitaler Arbeitsplätze einhergeht. Dazu kommen Geschäftsmodelle, die aus personenbezogenen Daten Werte schöpfen und Social-Media-Anwendungen, die durch das permanente Feedback zu den Handlungen einzelner auch grundlegende Rechte wie Privatheit oder Gleichbehandlung zumindest neu austarieren.

Die Erforschung dieser Effekte auf Einzelne, Organisationen und die Gesellschaft insgesamt, ist in den Sozialwissenschaften als Technologiefolgenabschätzung bereits im Gange und beeinflusst auch das Software Engineering.

Daher gibt es mehrere Schnittstellen des Software Engineering zu den Sozialwissenschaften, die auch deshalb relevant sind, weil viele Möglichkeiten der Digitalisierung und die daraus folgenden Verwendungsformen aktuell noch nicht umgesetzt sind, sondern höchstens angedacht werden. Dazu gehören sowohl Techniken der Speicherung und Verarbeitung personenbezogener Daten im Bereich Data Science als natürlich auch die Methoden der künstlichen Intelligenz, die im aktuellen Hype-Zyklus einige weitere Probleme lösen wird, aber auch dieses Mal alle nicht Probleme, zu deren Lösung Intelligenz notwendig ist, in den Griff bekommen dürfte. Software Engineering steht demnach zu den Sozialwissenschaften in folgenden Wechselwirkungen:

(1)     Software Engineering kann Szenarien liefern, die verständlich machen, welche Software-Funktionalitäten in der nächsten Zukunft realisierbar sind oder realisierbar werden können. Hier sind durchaus auch sowohl Szenarien interessant, die das Wohl einzelner oder das Gesamtwohl der Menschen verbessern, aber auch Szenarien beschreibbar, die nach dem aktuellen Werteverständnis zwar möglich, aber nicht wünschenswert sind. Dadurch können Untersuchungen über die langfristigen Konsequenzen sind und Diskussionen über das, was gesellschaftlich wünschenswert ist, ausgelöst und fakten-orientiert geführt werden.

(2)     Software Engineering nimmt Regulationen und Werteverständnisse auf und versucht Verfahren, Algorithmen und Architekturen zu entwickeln, um diese systematisch umzusetzen. Dieser Punkt ist gewissermaßen das Gegenstück zum vorherigen Punkt, wo das Software Engineering Szenarien liefert, hier nimmt es solche von den Sozialwissenschaften auf.

(3)     Software Engineering unterstützt Softwareentwicklungs-Ansätze mit Methoden und Werkzeugen, sodass auch Nicht-Softwaretechniker sie nutzen können. Als Beispiel können die bereits genannten Datenbanksysteme gelten oder auch das Enduser-Computing, das in verschiedensten Formen weitere Verbreitung findet. Dazu gehören einfache und intuitive Konfigurationssprachen, wie sie etwa im Bereich der Home-Automatisierung angeboten werden oder auch Wizards für bestimmte partikulare Probleme. Darüber hinaus trägt das Software Engineering auch zur Nutzbarmachung wesentlicher Algorithmen, insbesondere im Bereich des

Maschinellen Lernens von Zusammenhängen aus Datenmengen bei, wodurch den Sozialwissenschaften neue quantitative Methoden effektiv zur Verfügung stehen.

(4) Mit der Durchdringung aller Lebensbereiche mit softwarebasierten Produkten und Dienstleistungen rückt der Mensch als Nutzer, dessen User Experience es zu optimieren gilt, immer mehr in den Fokus der Entwicklung. Solche Usability Aspekte können nur mit sozialwissenschaftlichen Methoden, insbesondere aus der Psychologie, adäquat adressiert werden. Die Bedeutung der User Experience für den Erfolg von softwarebasierten Lösungen hat zusätzlich durch eine Vielzahl neuer Benutzerschnittstellen wie etwa für die Sprachein- und -ausgabe oder Augmented bzw. Virtual Reality stark zugenommen.

(5) Wie bereits besprochen haben Menschen und Organisationen einen entscheidenden Einfluss auf das Software Engineering. Die Entwicklung und Untersuchung von Software Engineering Artefakten bedarf deshalb der Anwendung von empirischen Forschungsmethoden und Theorien aus den Sozialwissenschaften wie der Psychologie oder Soziologie, um menschliche und organisatorische Kontextfaktoren geeignet in den Software Engineering Forschungsprozess integrieren zu können. Software Engineering hat in diesem Sinne nicht nur Züge von traditionellen Ingenieurwissenschaften wie Maschinenbau oder Elektrotechnik, sondern unterscheidet sich von diesen durch diesen ausgeprägteren sozialwissenschaftlichen Charakter.

Eigentlich war es schon immer eine Aufgabe des Software Engineering sicherzustellen, dass Software nachvollziehbar ist. Allerdings steigt die Komplexität der Software, die aus immer komplexeren Technologie-Stacks und immer mehr Anwendungskomponenten entsteht, weiter deutlich an. Es werden immer häufiger selbstlernende Komponenten aber auch traditionell programmierte intelligentere ("smarte") Komponenten zur Kontrolle immer kritischerer und mit immer mehr Sensorik behafteter und daher unsicherer Systeme eingebaut werden. Machine Learning-Komponenten komplizieren dies weiter. Die Verstehbarkeit von Software, also insbesondere die Nachvollziehbarkeit der Entscheidungen und Kontrolle, die Software trifft bzw. ausübt, wird daher in der nächsten Zeit deutlich wichtiger. Dabei ist Nachvollziehbarkeit während der Nutzung aber auch "post mortem", zum Beispiel zur Verbesserung der Software oder für gerichtliche Klärungen notwendig. Sollte das nicht nachhaltig gelingen werden roboterpsychologische Analysen für das Verhalten intelligenter Maschinen im Sinne Asimovs notwendig.

Während die anderen Ingenieurwissenschaften nutzen im Wesentlichen die Naturwissenschaften und die Mathematik als Grundlagenwissenschaften, sind dies für das Software Engineering vor allem die Mathematik aber auch die Sozialwissenschaften, weshalb eine Befruchtung in beide Richtungen (zwischen Software Engineering und Sozialwissenschaften) sinnvoll und notwendig ist.

# Was macht moderne SE Forschung aus und welche Themen müssen diskutiert werden

Um aktuelle und zukünftige Forschungsbeiträge des Software Engineering abzustecken und den Bezug zu anderen Wissenschaften zu präzisieren, ist es notwendig, die moderne Software Engineering-Forschung im Zeitalter der Digitalisierung zu charakterisieren.

Abbildung 1 zeigt die Kernbereiche der Software Engineering Forschung und ihre Wechselbeziehungen zu Anwendungsdomänen und zu Grundlagenwissenschaften. Im Sinne der oben genannten Kontextorientierung des Software Engineering [2] und der Durchdringung aller Lebensbereiche mit Software liefern Anwendungsdomänen wie die Produktion, die Logistik, das Gesundheitswesen, aber auch Wissenschaftsdisziplinen selbst (insbesondere auch die Informatik) den Kontext, d.h. Daten und Problemstellungen aus einem bestimmten Sachzusammenhang, für die Software Engineering-Forschung. In der Software Engineering-Forschung werden Artefakte (etwa Werkzeuge oder Entwicklungsmethoden) und Evidenz, d.h. empirisch oder formal fundierte Einsichten, bereitgestellt, um die Problemstellung im Anwendungskontext zufriedenstellend zu adressieren.

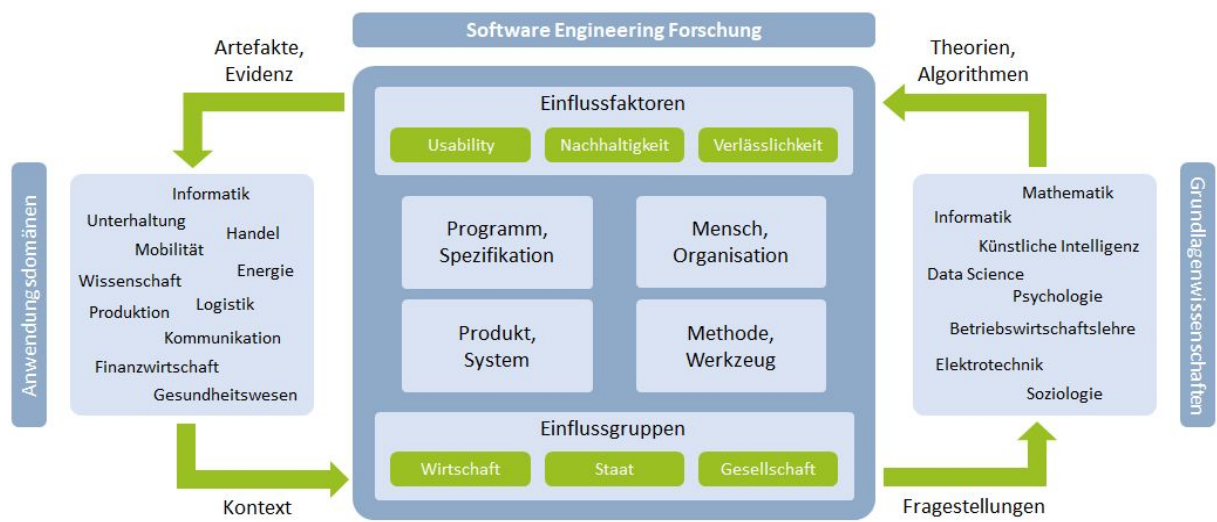

Abbildung 1: Software Engineering Forschung sowie Verbindung zu Anwendungsdomänen und Grundlagenwissenschaften

Als Ingenieursdisziplin greift die Software Engineering-Forschung dabei neben eigenen Forschungsansätzen auf Grundlagenwissenschaften wie Mathematik, Informatik, andere Ingenieurwissenschaften oder Sozialwissenschaften zurück. Diese liefern Algorithmen und Theorien und unterstützen dabei das Design von Lösungen und die Wissensgewinnung im Software Engineering  In den folgenden Abschnitten behandeln wir den Kern des Software Engineering also auch den Bezug zu Anwendungsdomänen und Grundlagenwissenschaften ausführlich.

# Kern der Software Engineering Forschung

Software Engineering und die zugehörige Forschung ist seit langem klar definiert, daran ändern auch aktuelle Diskussionen wie hier oder im Zuge von 50 Jahre Software Engineering [1] nichts. Die folgende bereits 1990 im IEEE Glossary of Software Engineering Terminology [5] vorgenommene Definition ist auch heute noch gültig und muss aus Sicht der Autoren nicht adaptiert werden:

> (1) The application of a systematic, disciplined, quantifiable approach to the development, operation, and maintenance of software; that is, the application of engineering to software.
> (2) The study of approaches as in (1).

Sie definiert das Software Engineering als Ingenieursdisziplin ("application of engineering to software") mit einer eigenen Methodik ("systematic, disciplined, quantifiable approach") angewendet auf alle Phasen des Lebenszyklus von Software ("development, operation, and maintenance of software"). Der zweiteilige Aufbau der Definition bringt auch die enge Verzahnung von Software Engineering (1) und Software Engineering Forschung (2) zum Ausdruck. Die Grenzen zwischen beiden verschwimmen bisweilen. In der Software Engineering Forschung ist auf alle Fälle die systematische Reflexion deutlich ausgeprägter und der angestrebte Nutzen sowie die Generalisierbarkeit der Ergebnisse jenseits einzelner Projekte deutlich größer. Eine scharfe Trennung kann jedoch nicht gezogen werden, da zum Beispiel die Entwicklung eines Frameworks, Werkzeugs oder Meta-Modells gleichzeitig sowohl Anwendungs- als auch Forschungs-Charakter haben kann.

Klassischerweise differenziert sich Software Engineering-Forschung selbst in die spezifische Betrachtung der unterschiedlicher Aktivitäten im Softwarelebenszyklus. Diese sind die bekannten Analyse, Design, Implementierung, Testen, Deployment und Wartung. Hinzu kommen Querschnittsaktivitäten wie Projektmanagement, Qualitätsmanagement, Modellierung und Modellierungssprachen oder Varianten-Management, die ebenfalls Teil der Software Engineering Forschung sind. Die zentrale Untersuchungsgegenstände in allen diesen Kern- und Querschnittsaktivitäten des Softwarelebenszyklus können vier Bereichen zugeordnet werden:

1. Das Programm (der Code oder dessen Ausführung) und die Spezifikation (das informelle Diagramm, das formale textuelle oder graphische Modell, die natürlichsprachliche Anforderung)
2. Der Mensch (als Entwickler, Anwender oder in einer der vielen weiteren Rollen) und die Organisation (Entwicklungsteam, Integration in ene Gesamtorganisation)
3. Das Produkt (etwa die softwaregetriebene Innovation, dessen Qualität, die Softwarelogistik) oder System (etwa die Integration von Software mit Hardware oder die Integration von Systemen)
4. Die Methode (das Verfahren, die menschliche Aktivität, der automatisierte Bearbeitungs-Algorithmus im Werkzeug, Prozessmodelle) oder das Werkzeug zur

Unterstützung aller Phasen des Softwarelebenszyklus (etwa Modellierungs- oder Testwerkzeuge)

Als Ingenieursdisziplin folgt das Software Engineering sinnvollerweise einem Design Science-Ansatz, bei dem Artefakte sich auf alle vorhin genannten Untersuchungsgegenstände des Software Engineering beziehen können. Der Begriff "Artefakt" ist dabei relativ breit angelegt und schließt Werkzeuge, zum Beispiel automatisierende oder unterstützende Algorithmen zum Beispiel zur Testfallgenerierung, Templates zur Dokumentation von Anforderungen oder eine Methodik für die Aufwandsschätzung von Softwareprojekten mit ein. Mit der Durchdringung diverser Produkte und Dienstleistungen mit Software, nimmt neben der systematischen Konstruktion von Artefakten die kreative Innovation eine immer bedeutendere Rolle ein und wird auch durch zunehmend durch Ansätze wie das Design Thinking oder Continuous Experimentation unterstützt. Werkzeuge spielen eine herausragende Rolle im Software Engineering, da sie die entwickelten Artefakte für die praktische Nutzung zugänglich machen. Erst die Existenz bestimmter Arten von Werkzeugen haben moderne Entwicklungsmethoden, wie zum Beispiel agile Entwicklung möglich gemacht. Extreme Programming wäre ohne leichtgewichtige Testinfrastruktur, schnelle Generatoren und Übersetzer, und Entwicklungsumgebungen mit integrierten Refactoring-Werkzeugen undenkbar. Der Komfort von spezifischen Editoren für Programmiersprachen (typspezifische Autovervollständigung, Highlighting, Refactoring im Editor) hat die Programmierweise der Entwickler massiv verändert. Hätten wir solche Werkzeuge auch in den früheren Phasen, wären auch Requirements Engineering, Architektur und Design wesentlich effizienter. Auch fehlen uns gute Werkzeuge zur Nachverfolgung von Anforderungen und Design-Entscheidungen, die man dringend benötigt, um den Impact von Änderungen zu antizipieren.

Software Engineering bzw. die Anwendung in konkreten Projekten beinhaltet typischerweise eine ausgeprägte Reflexion, die sowohl zur Optimierung innerhalb des Projekts als auch zur Weitergabe von projektübergreifendem Wissen genutzt werden. Daraus ergibt sich die Konsequenz, dass ein Teil der Software Engineering-Forschung notwendigerweise im industriellen Kontext vorgenommen werden muss und es deshalb des Einsatzes empirischer Methoden bedarf, um im industriellen Kontext Evidenz abzuleiten, etwa um Entscheidungen zu unterstützen, welches Prozessmodell in welcher Adaptierung in welchem Kontext am effektivsten funktioniert. Das Manifest für Agile Softwareentwicklung [7] etwa wurde 2001 auf einem Treffen von Praktikern formuliert und von diesen in die Praxis der Software-Entwicklung getragen. Die Aufgabe der Software Engineering-Forschung ist es, in diesem Zusammenhang die Methoden wissenschaftlich fundiert weiterzuentwickeln und deren praktischen Einsatz kritisch und empirisch fundiert zu hinterfragen und zu bewerten, um sie dadurch abhängig vom Anwendungskontext durch die Bereitstellung von Evidenz zu optimieren.

Aufgrund der Anwendungsorientierung und der durch Software getriebenen Digitalisierung spielen auch Einflussfaktoren und Einflussgruppen für die Software Engineering-Forschung eine große Rolle. Zu den Einflussfaktoren gehören die Verlässlichkeit ("Dependability"), die Nachhaltigkeit ("Sustainability") und die Benutzerfreundlichkeit ("Usability"). Verlässlichkeit ist traditionell stark vom Einsatz softwarebasierter Systeme in kritischen Domänen wie der

Luftfahrt beeinflusst und umfasst Attribute wie Zuverlässigkeit ("Reliability"), Sicherheit ("Safety"), Verfügbarkeit ("Availability" und damit letztlich auch Performanz), Wartbarkeit ("Maintainability") und Informationssicherheit ("Security"), aber zunehmend auch Nachvollziehbarkeit und Erklärbarkeit, speziell im Kontext moderner datengetriebener Anwendungen. Ausgehend vom ökologisch geprägten schonenden Umgang mit Ressourcen beinhaltet die Nachhaltigkeit auch ethische Aspekte wie die Fairness von Entscheidungsalgorithmen. Usability steht für die Benutzerfreundlichkeit oder Gebrauchstauglichkeit eines softwarebasierten Produkts oder eine Dienstleistung und gewinnt insbesonderen mit neuen Benutzerschnittstellen zunehmen an Bedeutung.
Als Einflussgruppen spielen die Wirtschaft, der Staat und die Gesellschaft eine große Rolle, etwa in dem sie den rechtlichen Rahmen für die Entwicklung und den Einsatz von Software abstecken, wie zum Beispiel durch die Datenschutzgrundverordnung in Bezug auf die Verarbeitung personenbezogener Daten.

## Relation des Software Engineerings zu ihren Grundlagenwissenschaften

Als Ingenieurswissenschaft ist das Software Engineering wie andere Ingenieurswissenschaften auch auf die Erkenntnisse von Grundlagenwissenschaften angewiesen. Für die anderen Ingenieurwissenschaften sind im Wesentlichen die Naturwissenschaften und die Mathematik die Grundlagenwissenschaften, im Software Engineering ist die Situation durch den immateriellen und den immanent humanen Charakter komplexer. Natürlich sind die Informatik, die etwa genetische Algorithmen für das Search-Based Software Engineering, oder die damit verwandte Künstliche Intelligenz bereitstellt und das Data Science als Datenwissenschaft wichtige Grundlagendisziplinen für das Software Engineering. Mit der zunehmenden Bedeutung der cyber-physikalischen Systeme sind auch andere technische Wissenschaften wie die Elektrotechnik oder der Maschinenbau bzw. die Physik als deren Grundlage wichtige Grundlagendisziplinen. Wie bereits erwähnt sind aber auch die Wirtschafts- und Sozialwissenschaften wie die Betriebswirtschaftslehre, die Soziologie oder die Psychologie wichtige Grundlagenwissenschaften für die Software Engineering Forschung. Wenn demnächst biologische Informationsverarbeitung stärker eine Rolle spielt, wie das heute Human Brain Projekte oder Modellierung biologischer Systeme mit Informatik-Methoden andeuten, werden auch Biologie und Chemie als Grundlage noch relevanter.

Der enge Bezug des Software Engineering zu Grundlagenwissenschaften ist bereits im Software Engineering Body of Knowledge (SWEBoK) [6] sichtbar, in dem eigene Kapitel Computing Foundations, Mathematical Foundations, Engineering Foundations im wesentlichen auf die Grundlagenwissenschaften Informatik, Mathematik und Ingenieurswissenschaften verweisen.

In der Software Engineering Forschung entstehen dabei bestimmte Fragestellungen, die mit Hilfe der Grundlagenwissenschaften adressiert werden. Diese stellen Algorithmen oder Theorien bereit, die formal-mathematischer, physikalischer oder sozialwissenschaftlicher Natur sein können, die im Forschungsprozess des Software Engineering in die Entwicklung und Evaluierung von Lösungen einfließen und die als Artefakte und Evidenz wieder in die Anwendungsdomänen zurückfließen können.

Je nach betrachtetem Untersuchungsgegenstand kann Software Engineering daher sehr Mathematik-nah und analytisch formal sein, zum Beispiel um formale Modelle oder Programme zu behandeln, aber auch sehr empirisch und experimentell, um etwa spezifische Hypothesen über den menschlichen Faktor bei der Entwicklung und der Nutzung zu untersuchen. Die dritte große Säule des Erkenntnisgewinns, die Simulation, ist im Software Engineering als Testen bereits intensiv im Einsatz und wird bei integrierter Qualitätssicherung von Software (beispielsweise durch Architektur- oder Verhaltenssimulation) in einem komplexen System wie etwa einer Smart City auch in anderen Formen noch intensiver eine Rolle spielen. Im Software Engineering fehlen aber immer noch Eigenschaften klassischer Ingenieursdisziplinen, wie der durchgängigen modellbasierten Simulation vor der Realisierung zur frühen Qualitätsbewertung. Abgesehen von einzelnen Ansätzen (wie Software-Architektursimulation) ist allerdings auch unklar, ob eine solche frühe Simulation auf Modellbasis zur Qualitätsbewertung von den klassischen Ingenieursdisziplinen auf das Software Engineering direkt übertragbar ist, da ja Software selbst im Gegensatz zu physischen Artefakten wie Brücken einen immanenten Modellcharakter hat.

Software und damit auch das Software Engineering ist seit jeher eine Integrationsdisziplin. Dies gilt natürlich für Produkte ganz besonders, die typischerweise aus Subsystemen und Komponenten integriert werden. Dies gilt aber auch für Entwicklungsmethoden, in denen je nach Art des Subsystems unterschiedlichste Methoden anderer Informatik-Teildisziplinen zum Einsatz kommen. Software-IngenieurInnen sind daher Generalisten. Demgegenüber gibt es praktisch für jeden Teilaspekt der Software jeweils eigenständige Spezialistenrollen. Dies ist so im Bereich der Netzwerke, der Betriebssysteme, der Datenbanken, der eingebetteten Systeme, der Cloud, des Übersetzerbaus, des Entwickelns von (domänenspezifischen) Modellierungssprachen, der Nutzer-Interaktion und natürlich auch der Sprach- und Bilderkennung mit statistischen Methoden oder maschinellem Lernen. Software Engineering greift deshalb nicht nur auf Grundlagendisziplinen wie die Informatik zu, sondern erweitert und bereichert diese auch. Aus diesem Grund sind die Informatik im Speziellen und Wissenschaft im Allgemeinen auch als Anwendungsdomänen in Abbildung 1 angeführt.

## Relation des Software Engineering zu ihren Anwendungsdomänen

Die großen Problemstellungen und Herausforderungen der Software-Entwicklung und damit auch der Software Engineering Forschung sind heute stark von der Anwendungsdomäne beeinflusst. Der Begriff Anwendungsdomäne kann dabei relativ breit gefasst werden und umfasst klassische Domänen wie Finanzen, Gesundheitswesen, Mobilität wie Automotive oder Avionik, Logistik, Unterhaltung, Handel oder Produktion. Darüber hinaus kann aber auch jede wissenschaftliche Disziplin wie die Informatik oder jede Sozial-, Ingenieurs- oder Naturwissenschaft, in denen heute größere Softwaresysteme im Einsatz sind, als Anwendungsdomäne verstanden werden.

Die Anwendungsdomänen stellen den Kontext, d.h. Daten und Problemstellungen aus einem bestimmten Sachzusammenhang, für die Software Engineering Forschung bereit. In der Software Engineering Forschung werden Artefakte (etwa Werkzeuge oder Entwicklungsmethoden) und/oder Evidenz, d.h. empirisch oder formal fundierte Einsichten, bereitgestellt, um die Problemstellung im Anwendungskontext zufriedenstellend zu adressieren.

Mittlerweile ist klar, dass zwar die Fragestellungen für die Softwareentwicklung in allen Domänen sehr verwandt sind, aber die Antworten häufig sehr unterschiedlich ausfallen müssen. Sichtbar wird dies unter anderem bereits an den sehr heterogenen Technologie-Stacks und den verwendeten Programmiersprachen. Dementsprechend gibt es für viele Domänen jeweils eigene Ausprägungen des Software Engineering und damit sinnvollerweise auch der Software Engineering Forschung. Domänenspezifisches Software Engineering bedeutet immer auch intensive Zusammenarbeit mit den Domänenexperten.

# Wo geht die Reise hin

## Digitalisierung des Software und des Systems Engineeering

Die aktuell stattfindende Digitalisierung praktisch aller Domänen erfordert domänenspezifische, effektive Verfahren Software schnell zu erstellen und wiederverwendbar weiterzuentwickeln. Zu viele Domänen sind heute noch getrieben von den dort üblichen Entwicklungsmethoden, die typischerweise in Konflikt zur Software-Entwicklungsmethodik stehen und daher in Projekten regelmäßig Friktionen produzieren. Dies gilt insbesondere für die klassischen Ingenieursdomänen wie Maschinenbau, Bauingenieurwesen oder Elektrotechnik. Solange die Maschine das dominante Element des Produkts ist, muss natürlich der Entwicklungsprozess entlang der Maschine und ihrer Komponenten definiert werden und daher typischerweise geometrische Dekomposition anbieten. Da aber immer mehr die Software-Funktionalität ansteigt und die Software der Innovationstreiber wird, ist es sinnvoll über eine Neuverhandlung der Entwicklungsprozesse nachzudenken, die eine funktionale Dekomposition entlang der Software-Funktionalität stärker in den Fokus nehmen. Die heute üblichen Software-Entwicklungsmethoden, wie etwa das V-Modell oder agile Methoden, sind jedoch auch nicht direkt auf andere Disziplinen übertragbar, obwohl die beobachtbare und durchaus berechtigte Hoffnung besteht, dass eine Generalisierung von Software Engineering Methoden für andere Domänen funktionieren dürfte.  Es ist eine herausragende Aufgabe des Software Engineering in Kooperation mit den traditionellen Ingenieursdisziplinen integrierte, ganzheitliche Entwicklungsmethoden und darauf aufbauende Werkzeuge zu entwickeln. Allerdings ist die Entwicklung eines Werkzeugs, das von Praktikern zur Anwendung genommen wird, heute keineswegs trivial. Explizite Werkzeugforschung ist notwendig, um den Bau industriell nutzbarer Werkzeuge überhaupt zu ermöglichen, denn moderne und damit oft auch intelligente Werkzeuge könnten die Softwareentwicklung weit jenseits des heutigen Standes unterstützen. Dies gilt allerdings nur, wenn solche Werkzeuge einerseits eine gewisse Reife besitzen und andererseits durch explizit offene Schnittstellen und Modularisierungstechniken Erweiterbarkeit sowie projektspezifische Anpassbarkeit ermöglichen. Das Systems Engineering, und speziell der Einsatz von Modellierungs- und Simulationstechniken, speziell in der Qualitätssicherung, ist dabei eine wesentliche Herausforderungen, um die Zeit zur Marktreife sowie die Entwicklungskosten innovativ weiterentwickelter Produkte deutlich zu verkürzen. Dies ist gerade in Hochlohn-Ländern essentiell, um die Innovationsfähigkeit nicht zu verlieren.

## Das Verhältnis des Software Engineering zu Daten

Natürlich erhalten auch Daten, die immer mehr domänenspezifische Information in sich tragen, in der Funktionalität von Software eine immer größere Rolle. Bisher wurden Daten zwar als Gegenstand der Software-Funktionen verarbeitet, sie haben aber nicht selbst wesentlich die Funktionalität bestimmt wie das zunehmend in modernen datenintensiven Systemen der Fall sein wird. Das ändert sich nun durch Verfahren des Maschinellen Lernens. Dadurch entfällt die Möglichkeit, die Software zur Entwicklungszeit vollumfänglich zu analysieren oder zu testen. Deshalb sind Verfahren zur Absicherung selbstlernender, hoch-adaptiver Software notwendig, die klären, unter welchen Bedingungen Systeme noch zum Entwurfszeitpunkt analysierbar sind, Verfahren zur Laufzeitabsicherung und Verfahren zur a-posteriori-Fehleranalyse. Darüber hinaus spielen Daten noch eine weitere Rolle als „first class entities" bei der Modellierung von Privatheitsanforderungen und allgemeiner Zugriffs-Regelungen.

Die zunehmende Verfügbarkeit von Daten bietet auch für die Software Engineering Forschung eine große Chance, um Hypothesen empirisch zu prüfen. Allerdings birgt die beinahe beliebige Verfügbarkeit bestimmter Arten von Daten etwa aus Open Source Repositories und die sehr schwierige Erhebung anderer Datenarten etwa im Industrie 4.0 Kontext im Software Engineering auch Gefahren für den Forschungsprozess. Zu häufig werden Daten etwa aus Software-Repositorien oder durch Umfragen generiert und publiziert ohne diese durch eine Theorie, eine genaue Kontextdefinition oder eine Vision zu fundieren, wodurch der Nutzen dieser Arbeiten für die Disziplin des Software Engineering unklar bleibt. Mit Blick auf die Naturwissenschaften bringt der Philosoph Karl R. Popper mit seiner Aussage „Alle Erkenntnis ist theoriegetränkt, auch unsere Beobachtungen" [8] zum Ausdruck, dass Experimente zur Erhebung von Daten gemäß einer zu prüfenden Hypothese aufgebaut sein sollen und diese Hypothese aus einer Theorie folgt. Eine Datenerhebung ohne zugrundeliegende Theorie mag zwar der initialen Phase eines Forschungsprozesses als explorative Studie durchaus berechtigt sein, um daraus Hypothesen abzuleiten. Diese sind dann aber durch weitere empirische Studien zu evaluieren. Dazu kommt, dass die Verfügbarkeit von vielen Open Source Repositorien der Blick darauf verstellt, dass Software nicht nur auf Code-Artefakten beruht.

## Qualität: Verlässlichkeit, Nachvollziehbarkeit, Erklärbarkeit, etc.

Verlässlichkeit (Dependability), also der Grad, in dem darauf vertraut werden kann, dass ein System zugesicherte Leistungen erbringt bzw. zugesicherte Eigenschaften hat, gewinnt in Bezug auf Zuverlässigkeit, Security und Safety von Systemen eine immer größere Bedeutung. Gleichzeitig wird es aber für stark vernetzte, komplexe und autonome Systeme immer schwieriger, Verlässlichkeit und darüber hinaus Nachvollziehbarkeit und Erklärbarkeit von Systemen (etwa auf zur gesellschaftlich notwendigen Klärung von Haftungsfragen bei schädigendem Verhalten autonomer Systeme, was als Nachhaltigkeitsaspekt verstanden werden kann) sicherzustellen. Ein vielversprechender Ansatz, um dieses Problem zu adressieren, sind Abstraktionstechniken basierend auf Modellen, die helfen, die Komplexität, Vernetztheit, Autonomie und Nachvollziehbarkeit beherrschbar zu machen. Modellbasierte Ansätze sollen dabei sowohl generierende Modelle, beispielsweise für das modellbasierte Testen, als auch prädiktive Modelle, basierend auf maschinellem Lernen, integrieren. Diese Abstraktionstechniken sollten es dann ermöglichen, etwa relevante Eigenschaften des Verhaltens von High-Frequency-Trading-Techniken in der Finanzwirtschaft abzubilden, um das Zusammenwirken von mehreren dieser Algorithmen auf einem offenen Markt mit zugehörigen Analysetechniken zu untersuchen. Eine Herausforderung ist hierbei, wie man zum einen von den konkreten Details der Algorithmen soweit abstrahieren kann, dass Geschäftsgeheimnisse gewahrt werden und die Analyse noch skaliert, aber zum anderen noch aussagekräftige Analyseergebnisse gewonnen werden können, die ein Mensch die Ergebnisse noch nachvollziehen kann. Das Beispiel High-Frequency-Trading aus der Finanzwirtschaft zeigt auch nochmals die Rolle der Domäne.

Anwendungsdomänen entwickeln immer eine Eigendynamik und erfordern jeweils eigene Ausprägungen des Software Engineering und damit sinnvollerweise auch der Software Engineering Forschung.

## Konkrete Fragestellungen des Software Engineering

Aus den bisherigen Betrachtungen ergeben sich viele konkrete Fragestellungen, welche die Software Engineering Forschung adressieren soll, um das Software Engineering und deren Forschung voranzubringen. Beispiele für konkrete Fragestellungen sind etwa:

- Nachvollziehbarkeit und Erklärbarkeit von Software: Wie können auch bei nichtdeterministischen datenintensiven Softwaresystemen, Ergebnisse plausibel erklärt werden?
- Security als ganzheitliches Thema: Wie können Security Aspekte geeignet im Softwarelebenszyklus berücksichtigt werden. Wie kann etwa eine Brücke aussehen zwischen abstrakte Sicherheitsmodelle mit formalen Nachweisen und realer Software, die in dynamischen und offenen Umgebungen evolviert?
- Software Engineering und Nachhaltigkeit: Welche Software Engineering Methoden werden benötigt, um Ressourcenmanagement zu steuern und verstehen zu können?
- Forschungsmethodik im Software Engineering: Was können empirische und formale Methoden im Forschungsprozess leisten? Welche Rolle nehmen Kontextfaktoren (insbesondere auch menschliche und organisatorische Faktoren) ein und welche Theorien benötigt das Software Engineering und wie sehen diese aus? Welche Instrumente können Design Science im Software Engineering am besten unterstützen?
- Automatisierung durch Werkzeuge: Das in der Softwareentwicklung noch sehr viel mehr automatisiert werden kann ist unstrittig. Dadurch steigen Effizienz der EntwicklerInnen und Qualität der Produkte. Wie kann das schlummernde Potenzial besserer und automatisierter Unterstützung gehoben werden? Welche Werkzeuge sind besonders hilfreich? Wie können die Werkzeuge gestaltet werden, damit die Einstiegshürde möglichst gering ausfällt? Wie können Forschungsprototypen zu kommerziell einsetzbaren Werkzeugen weiterentwickelt werden?
- Qualitätssicherung für datenintensive Systeme: Mit welchen Methoden kann die Qualität von Softwaresystemen, deren Verhalten durch Daten und die Umgebung bestimmt wird und zur Entwicklungszeit noch nicht feststeht, geeignet sichergestellt werden? Wie können Simulation, modellbasierte Softwareentwicklung, Runtime-Monitoring sowie statische Analyse und dynamisches Testen dabei geeignet integriert werden?
- Software in Systems of Systems: Wie geht man in Entwicklungsprozessen damit um, dass es keine zentrale Organisation mehr für Anforderungen und Entwicklungskoordination gibt? Was bedeutet Konsistenz in solchen Systemen und wie kann man sie wiederherstellen? Wie bringt man unterschiedliche Lebenszyklus-Modelle und Update-Geschwindigkeiten in Einklang? Wie integriert man die lebenslange Verbindung (Messung, Update) zwischen dem produzierten und ausgelieferten Produkt sowie seinem Nutzer und Hersteller?
- Umgang mit immer komplexeren Entwicklungs- und Laufzeitumgebungen: Wie kann man bei steigender Funktionalität, EntwicklerInnen das methodisches Wissen an die Hand geben, um solche Elemente einzusetzen bzw. zu entwickeln und zu pflegen?

## Herausforderungen des Software Engineerings

Das Software Engineering ist keine Natur-, sondern eine Ingenieurwissenschaft mit starkem Fokus auf die vier Säulen aus Abbildung 1:

(1) Programm und Spezifikation,
(2) Mensch und Organisation,
(3) Produkt und System, sowie
(4) Methode und Werkzeug.

Diese Säulen heben das Software Engineering von den traditionellen Ingenieurswissenschaften ab. Gerade wegen der am Entwicklungsprozess beteiligten Menschen treten anstelle von naturwissenschaftlichen Theorien oft Hypothesen, die sich aus Methodenvorschlägen ergeben und insbesondere auch menschliche und organisatorische Kontextfaktoren berücksichtigen. Diese Hypothesen über die Eigenschaften einer neu vorgeschlagenen Methode müssen dann empirisch geprüft werden etwa durch kontrollierte Experimente oder Fallstudien. Um eine neu vorgeschlagene Methode im Software Engineering sinnvoll anwendbar zu machen und der empirischen Überprüfung zu unterziehen, ist es meistens notwendig diese ganz oder teilweise in Werkzeugen zu implementieren. Nur so kann der Automatisierungsgrad der Methode, der durch ein Werkzeug ermöglicht wird, in Fallstudien eingebracht und empirisch validiert werden. Werkzeugbau ist daher eine notwendige und wichtige die Forschung begleitende Tätigkeit. Dass so zumindest auch Prototypen für Werkzeuge in der Softwareentwicklung entstehen ist für die Industrialisierung neuer Software Engineering Methoden zumindest hilfreich und müsste auch politisch mehr unterstützt werden.

Es ist eine zentrale Aufgabe des Software Engineering neue Visionen zu entwickeln und diese in Methoden und Werkzeuge zu gießen und begleitend deren Nützlichkeit, Effektivität und Effizienz empirisch zu evaluieren. Nur so ist es möglich die tragende Rolle des Software Engineering für Wirtschaft und Gesellschaft im Zeitalter der Digitalisierung zu halten und weiter auszubauen. Gemäß der Aussage des Automobilfabrikanten Henry Fords „Wenn ich meine Kunden gefragt hätte, was sie wünschen, hätten sie gesagt: schnellere Pferde" ist es notwendig Visionen in praxistauglichen Methoden und Werkzeuge umzusetzen.

Dabei ist die Methodenweiterentwicklung gerade im Zusammenhang mit dem beschriebenen Wandel der Rolle des Software Engineerings und ihrer stärkeren Verzahnung mit anderen Disziplinen (insbesondere dem Systems Engineering) außerordentlich spannend. War bisher die Entwicklung von Software ein Teilschritt im übergeordneten Systems Engineering, so wandelt sich dieses Verhältnis bei der enorm zunehmende Rolle von Software bei fast jeder technischen Innovation. Diese zunehmend zentrale Rolle der Software wird sich aus Sicht der Autoren auch in den Methoden zur Systementwicklung niederschlagen - diese werden software-zentrierter. Gemäß dieser neuen Rolle von Software bei Innovationen, wird das Produkt durch seine Software getrieben, die elektrischen oder mechanischen Komponenten sind entweder Commodity, oder werden entwickelt entsprechend der Anforderungen aus der Software. Dies bietet eine einzigartige Chance für das Software Engineering als wichtiger Treiber im Systems Engineering aufzutreten. Diese Chance kann man nur genutzt, wenn sich Software Engineering Forscher ihrer Rolle als Ingenieurwissenschaftler und Innovatoren bewusst sind, deren Forschung Visionen für neuen Methoden hervorbringt.

## Danksagung

## Referenzen